\documentclass[12pt,a4paper]{article}

\usepackage[utf8]{inputenc}
\usepackage[russian]{babel}
\usepackage{fullpage}

\usepackage{amsmath}
\usepackage{amssymb}
\usepackage{amsthm}
\usepackage{fixmath}

\usepackage{tikz}
\usepackage{tikz-inet}

\tikzset{every node/.style = {node distance=0em}}

\title{Модели вычислений,\\
основанные на взаимодействии}
\author{(неофициальный перевод)}

\theoremstyle{definition}
\newtheorem{theorem}{Утверждение}
\newtheorem{definition}[theorem]{Определение}
\newtheorem{example}[theorem]{Пример}

\begin{document}
\maketitle

\begin{abstract}
Данный текст представляет собой перевод на русский язык седьмой главы учебника по моделям вычислений \cite{tutor}.
В этой главе изучаются сети взаимодействия~---~модель вычислений, которую можно рассматривать как представителя класса моделей, основанных на идее <<вычислений через взаимодействие>>.
Сети взаимодействия являются \textit{графической} моделью вычислений, введенной в~\cite{inet} как обобщение структур доказательств линейной логики.
Данный формализм используется для описания алгоритмов и анализа их сложности, а также как язык низкого уровня, в который можно компилировать другие языки программирования.
Особенно интересны сети взаимодействия тем, что они могут быть реализованы с разумной эффективностью.

Системы сетей взаимодействия определяются двумя множествами: набором \textit{агентов} и набором \textit{правил взаимодействия}.
Агенты играют роль логических символов, а правила взаимодействия определяют их смысл.
Также возможна аналогия с электрическими цепями, где агенты играют роль узлов цепи, а связи между агентами служат ветвями, соединяющими узлы.
Наконец, мы можем просто считать агенты некоторыми объектами, выполняющими вычисления, при этом правила взаимодействия описывают их поведение.

Здесь дается общее представление о парадигме взаимодействия, приводятся примеры использования сетей взаимодействия для представления алгоритмов, а также показывается, как другие модели вычислений представляются в виде сетей взаимодействия.
\end{abstract}

\section{Парадигма взаимодействия}

Системы сетей взаимодействия определяются множеством $\Sigma$ символов, из которых сеть может состоять, и множеством $\mathcal R$ так называемых \textit{правил взаимодействия}~---~правил перезаписи для сетей взаимодействия, удовлетворяющих определенным условиям, обсуждаемым ниже.

Каждый символ $\alpha \in \Sigma$ будет иметь некоторую (фиксированную) \textit{арность}~---~натуральное число (с нулем).
Условимся, что вне зависимости от рассматриваемого множества $\Sigma$ функция $\text{Ar}: \Sigma \rightarrow \mathbb N$ будет возвращать арность, связанную с данным символом из $\Sigma$.

Более точно сети определяются следующим образом.

\begin{definition}[cеть]
Для данного множества $\Sigma$ \textit{сеть} $N$ является ориентированным графом (граф не обязан быть связным), в котором вершины помечены символами из $\Sigma$.
Помеченная таким образом вершина называется \textit{агентом}; дуга между двумя агентами называется \textit{связью}; т.~е. сети состоят из агентов и связей.

Точки присоединения связей к агентам называются \textit{портами}.
Если агент $\alpha$ имеет арность $n$, то каждая вершина, помеченная $\alpha$ должна иметь $n + 1$ портов: один порт называется \textit{главным}, а остальные $n$~---~\textit{дополнительными} портами.
Количество дополнительных портов каждого агента равно арности соответствующего символа.
\end{definition}

Принимая во внимание ориентированность графа, пронумеруем порты против часовой стрелки, начиная с главного порта.
Если $\text{Ar}(\alpha) = n$, то агент можно изобразить следующим образом.
$$
\begin{tikzpicture}[baseline=(i.base)]
\inetcell(a){$\phantom X\alpha\phantom X$}[R]
\inetwirefree(a.pal)
\inetwirefree(a.left pax)
\inetwirefree(a.right pax)
\node (0) [right=of a.above pal] {$x_0$};
\node (1) [left=of a.above left pax] {$x_1$};
\node (i) [left=of a.above middle pax] {$\vdots$};
\node (n) [left=of a.above right pax] {$x_n$};
\end{tikzpicture}
$$
Если $\text{Ar}(\alpha) = 0$, то агент $\alpha$ не будет иметь дополнительных портов, но каждый агент всегда имеет главный порт.

В сети взаимодействия дуги соединяют агентов вместе так, что на каждый порт приходится не более одной дуги, при этом дуги также могут соединять порты одного и того же агента.
Порты агента, которые не связаны с каким-либо другим портом в сети, называются \textit{свободными}.
Стоит обозначить два особых вида сетей.
Сеть может состоять исключительно из связей (без агентов); такие сети мы будем называть \textit{системами связей}, причем концы дуг в этом случае мы также будем называть портами.
Если в системе связей имеется  $n$ дуг, то она содержит $2n$ свободных портов.
Если же сеть не содержит ни агентов, ни связей, то мы будем называть ее \textit{пустой сетью}.
Наконец, \textit{интерфейсом} сети назовем множество ее свободных портов.

\begin{definition}[правило взаимодействия]
Пара агентов $(\alpha, \beta) \in \Sigma \times \Sigma$, соединенных вместе своими главными портами называется \textit{активной парой}; мы будем обозначать ее как $\alpha \bowtie \beta$.
Активные пары являются аналогами редексов в $\lambda$-исчислении.

\textit{Правило взаимодействия} $\alpha \bowtie \beta \rightarrow N$ в $\mathcal R$ состоит из активной пары слева и сети $N$~---~справа;
при этом правила взаимодействия должны удовлетворять следующим двум сильным условиям.
\begin{enumerate}
\item В правиле взаимодействия левая и правая части должны иметь один и тот же интерфейс, то есть взаимодействие сохраняет все свободные порты.
Это свойство можно проиллюстрировать на схеме.
$$
\begin{tikzpicture}[baseline=(yi.base)]
\matrix[column sep=1em]{
\inetcell(a){$\phantom X\alpha\phantom X$}[R] &
\inetcell(b){$\phantom X\beta\phantom X$}[L] \\ };
\inetwirefree(a.left pax)
\inetwirefree(a.right pax)
\inetwire(a.pal)(b.pal)
\inetwirefree(b.left pax)
\inetwirefree(b.right pax)
\node (x1) [left=of a.above left pax] {$x_1$};
\node (xi) [left=of a.above middle pax] {$\vdots$};
\node (xn) [left=of a.above right pax] {$x_n$};
\node (y1) [right=of b.above left pax] {$y_1$};
\node (yi) [right=of b.above middle pax] {$\vdots$};
\node (yn) [right=of b.above right pax] {$y_n$};
\end{tikzpicture}
\rightarrow
\begin{tikzpicture}[baseline=(xi.base)]
\matrix[column sep=2em]{
\node (x1) {$x_1$}; &
\node (t) {$\phantom x$}; &
\node (yn) {$y_n$}; \\
\node (xi) {$\vdots$}; &
\node (n) {$N$}; &
\node (yi) {$\vdots$}; \\
\node (xn) {$x_n$}; &
\node (b) {$\phantom x$}; &
\node (y1) {$y_1$}; \\ };
\inetbox{(b) (t)}(box)
\inetwirecoords(x1)(intersection cs:
first line={(box.north west)--(box.south west)},
second line={(x1)--(yn)})
\inetwirecoords(xn)(intersection cs:
first line={(box.north west)--(box.south west)},
second line={(xn)--(y1)})
\inetwirecoords(yn)(intersection cs:
first line={(box.north east)--(box.south east)},
second line={(x1)--(yn)})
\inetwirecoords(y1)(intersection cs:
first line={(box.north east)--(box.south east)},
second line={(xn)--(y1)})
\end{tikzpicture}
$$
Заметим, что сеть $N$ сама может содержать агенты $\alpha$ и $\beta$.
Также $N$ может быть системой связей, если только суммарное количество свободных портов в активной паре было четным.
В случае активной пары вовсе без свободных портов $N$ может (но не обязана) быть пустой сетью.

\item Множество $\mathcal R$ может содержать не более одного правила для каждой неупорядоченной пары агентов, то есть только одно правило для $\alpha \bowtie \beta$, которое совпадает с правилом для $\beta \bowtie \alpha$.
\end{enumerate}
\end{definition}

Правила взаимодействия порождают отношение редукции на сетях.

\begin{definition}[редукция]
Шаг редукции по правилу $\alpha \bowtie \beta \rightarrow N$ заменяет одну из активных пар $\alpha \bowtie \beta$ сетью $N$.
Более точно, отношение \textit{редукции} $W \rightarrow W'$ между сетями $W$ и $W'$ имеет место, если сеть $W$ содержит активную пару $\alpha \bowtie \beta$ и множество $\mathcal R$ содержит правило взаимодействия $\alpha \bowtie \beta \rightarrow N$, такую, что сеть $W'$ получается заменой активной пары $\alpha \bowtie \beta$ в сети $W$ сетью $N$ (после такой замены не появляется <<висящих>> дуг, так как $N$ имеет тот же интерфейс, что и $\alpha \bowtie \beta$).

Рефлексивное транзитивное замыкание отношения редукции обозначим <<$\rightarrow^*$>>.
Другими словами, мы пишем $N \rightarrow N'$, если $N'$ может быть получена из $N$ редукцией одной активной пары, а $N \rightarrow^* N'$ означает, что существует последовательность из нуля или более шагов взаимодействия, приводящих сеть $N$ к виду $N'$.
\end{definition}

Вообще говоря, не требуется, чтобы имелось правило взаимодействия для любой пары агентов, но если сеть содержит активную пару, для которой нет правила взаимодействия, то эта пара не будет редуцироваться (она будет \textit{заблокирована}).

Заметим, что интерфейс сети \textit{упорядочен}.
Используя это свойство, мы может избавиться от обозначений для свободных портов.
Например, правило взаимодействия
$$
\begin{tikzpicture}[baseline=(p.base)]
\matrix[column sep=1em]{
\inetcell(a){$\alpha$}[R] &
\inetcell(b){$\beta$}[L] \\ };
\inetwirefree(a.left pax)
\inetwirefree(a.right pax)
\inetwire(a.pal)(b.pal)
\inetwirefree(b.left pax)
\inetwirefree(b.right pax)
\node (x1) [left=of a.above left pax] {$x_1$};
\node (xn) [left=of a.above right pax] {$x_2$};
\node (y1) [right=of b.above left pax] {$y_1$};
\node (yn) [right=of b.above right pax] {$y_2$};
\node (p) [right=of b.middle pax] {$\phantom x$};
\end{tikzpicture}
\rightarrow
\begin{tikzpicture}[baseline=(p.base)]
\matrix[column sep=2em]{
\node (x1) {$x_1$}; & & \node (y2) {$y_2$}; \\
& \node (p) {$\phantom x$}; & \\
\node (x2) {$x_2$}; & & \node (y1) {$y_1$}; \\ };
\inetwirecoords(x1)(y1)
\inetwirecoords(x2)(y2)
\end{tikzpicture}
$$
соединяет $x_1$ с $y_1$ и $x_2$ с $y_2$, что эквивалентно следующей схеме.
$$
\begin{tikzpicture}[baseline=(p.base)]
\matrix[column sep=1em]{
\inetcell(a){$\alpha$}[R] &
\inetcell(b){$\beta$}[L] \\ };
\inetwirefree(a.left pax)
\inetwirefree(a.right pax)
\inetwire(a.pal)(b.pal)
\inetwirefree(b.left pax)
\inetwirefree(b.right pax)
\node (x1) [left=of a.above left pax] {$x_1$};
\node (xn) [left=of a.above right pax] {$x_2$};
\node (y1) [right=of b.above left pax] {$y_1$};
\node (yn) [right=of b.above right pax] {$y_2$};
\node (p) [right=of b.middle pax] {$\phantom x$};
\end{tikzpicture}
\rightarrow
\begin{tikzpicture}[baseline=(p.base)]
\matrix[column sep=2em]{
\node (x1) {$x_1$}; & & \node (y1) {$y_1$}; \\
& \node (p) {$\phantom x$}; & \\
\node (x2) {$x_2$}; & & \node (y2) {$y_2$}; \\ };
\inetwirecoords(x1)(y1)
\inetwirecoords(x2)(y2)
\end{tikzpicture}
$$
Однако, в последней записи существены метки на концах свободных портов, так как мы изменили их порядок.
В дальнейшем мы всегда будем сохранять порядок свободных портов при записи правил взаимодействия, чтобы избежать необходимости в метках для интерфейсов.

Сеть взаимодействия находится в \textit{полной нормальной форме} (обычно мы будем называть ее просто \textit{нормальной формой}), если в ней нет активных пар.
Обозначим через $N \downarrow N'$ тот факт, что существует конечная последовательность взаимодействий $N \rightarrow^* N'$, такая, что $N'$ находится в нормальной форме.
Если $N \downarrow N'$, то сеть $N$ будем называть \textit{нормализуемой}; если же все последовательности взаимодействий, начинающиеся с сети $N$, конечны, то $N$~---~\textit{сильно нормализуема}.

Одним из прямых следствий определения систем взаимодействия, в частности ограничений, накладываемых на правила взаимодействия, является следующее полезное свойство редукции на сетях взаимодействия.
Если возможны две различные редукции сети $N$ (т.~е. $N \rightarrow N_1$ и $N \rightarrow N_2$), то существует сеть $M$, такая, что обе сети $N_1$ и $N_2$ могут быть редуцированы к $M$ за \textit{один шаг}: $N_1 \rightarrow M$ и $N_2 \rightarrow M$.
Это свойство, иногда называемое \textit{сильной конфлюэнтностью} или свойством ромба, сильнее конфлюэнтности; из него следует конфлюэнтность.
Таким образом, мы получаем следующий результат.

\begin{theorem}
Пусть $N$~---~сеть в системе взаимодействия $(\Sigma, \mathcal R)$.

\begin{enumerate}
\item Если $N \downarrow N'$, то сеть $N$ сильно нормализуема, т.~е. любые последовательности редукций, начинающиеся с $N$, завершаются.

\item Если $N \downarrow N'$ и $N \downarrow N''$, то $N' = N''$ (нормальные формы уникальны).
\end{enumerate}
\end{theorem}

Представим две простые, но важные операции на сетях взаимодействия.

\begin{example}
\label{erasedup}
Чаще других рассматриваются \textit{удаляющий} агент $\epsilon$, который приводит к удалению любых агентов, с которыми он взаимодействует, и \textit{дублирующий} агент $\delta$, который, наоборот, дублирует любого агента, с которым он взаимодействует.
Точнее, правила взаимодействия для этих двух агентов выглядят следующим образом.
$$
\begin{tikzpicture}[baseline=(a)]
\matrix[row sep=1em]{
\inetcell(e){$\epsilon$}[D] \\
\inetcell(a){$\phantom x\alpha\phantom x$}[U] \\ };
\node (i) [below=of a.middle pax] {$\dots$};
\inetwirefree(a.left pax)
\inetwirefree(a.right pax)
\inetwire(e.pal)(a.pal)
\end{tikzpicture}
\rightarrow
\begin{tikzpicture}[baseline=(i.base)]
\matrix{
\inetcell(1){$\epsilon$} &
\node (i) {$\dots$}; &
\inetcell(n){$\epsilon$} \\ };
\inetwirefree(1.pal)
\inetwirefree(n.pal)
\end{tikzpicture}
\qquad
\begin{tikzpicture}[baseline=(a)]
\matrix[row sep=1em]{
\inetcell(d){$\delta$}[D] \\
\inetcell(a){$\phantom x\alpha\phantom x$}[U] \\ };
\node (i) [below=of a.middle pax] {$\dots$};
\inetwirefree(a.left pax)
\inetwirefree(a.right pax)
\inetwirefree(d.left pax)
\inetwirefree(d.right pax)
\inetwire(d.pal)(a.pal)
\end{tikzpicture}
\rightarrow
\begin{tikzpicture}[baseline=(i.base)]
\matrix[row sep=2em]{
\inetcell(l){$\phantom x\alpha\phantom x$}[U] & &
\inetcell(r){$\phantom x\alpha\phantom x$}[U] \\
\inetcell(1){$\delta$} &
\node (i) {$\dots$}; &
\inetcell(n){$\delta$} \\ };
\node (li) [below=of l.middle pax] {$\dots$};
\node (ri) [below=of r.middle pax] {$\dots$};
\inetwire(l.left pax)(1.right pax)
\inetwire(l.right pax)(n.right pax)
\inetwire(r.left pax)(1.left pax)
\inetwire(r.right pax)(n.left pax)
\inetwirefree(1.pal)
\inetwirefree(n.pal)
\inetwirefree(l.pal)
\inetwirefree(r.pal)
\end{tikzpicture}
$$
На каждой из двух схем выше символом $\alpha$ обозначается произвольный агент из $\Sigma$, поэтому они представляют собой одновременно несколько правил взаимодействия.
Первая схема определяет взаимодействие активной пары $\epsilon \bowtie \alpha$, а вторая~---~$\delta \bowtie \alpha$.

Первое правило определяет взаимодействие между произвольным агентом $\alpha$ и удаляющим агентом $\epsilon$ как удаление $\alpha$ и создание удаляющих агентов на всех свободных портах $\alpha$.
Заметим, что если $\alpha$ имеет нулевую арность, то правая часть правила будет пустой сетью; в этом случае процесс удаления завершается.
Частным случаем является взаимодействие двух удаляющих агентов, когда $\alpha$ есть сам $\epsilon$.
Эти правила обеспечивают механизм так называемой <<сборки мусора>> для сетей взаимодействия.

Во втором правиле мы видим, что агент $\alpha$ дублируется, при этом на всех свободных портах $\alpha$ создаются дублирующие агенты $\delta$, которые таким образом могут продолжить дублировать остальную часть сети.
\end{example}

\section{Числа и арифметические операции}
\label{numbers}

Натуральные числа могут быть представлены с помощью нуля $0$ и функции следования $S$.
Например, числу $3$ соответствует выражение $S(S(S(0)))$.
Теперь рассмотрим определение стандартной операции сложения:
\begin{align*}
\text{Add}(0, y) &= y; \\
\text{Add}(S(x), y) &= S(\text{Add}(x, y)).
\end{align*}
Действительно, сложение любого числа $y$ с нулем дает в результате $y$, а для того чтобы прибавить $x + 1$ к $y$, можно вычислить $x + y$ и увеличить результат на $1$.

Чтобы представить этот алгоритм на языке сетей взаимодействия, введем три агента, соответствующих $\text{Add}$, $S$ и $0$.
$$
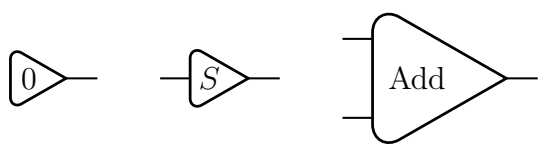
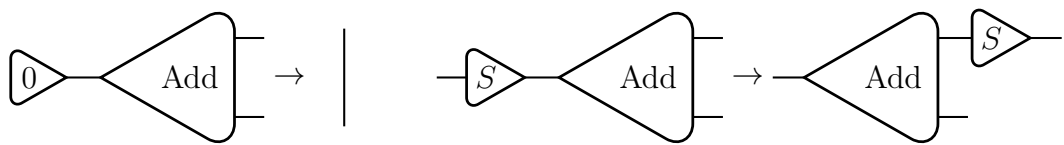
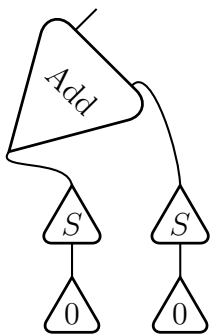
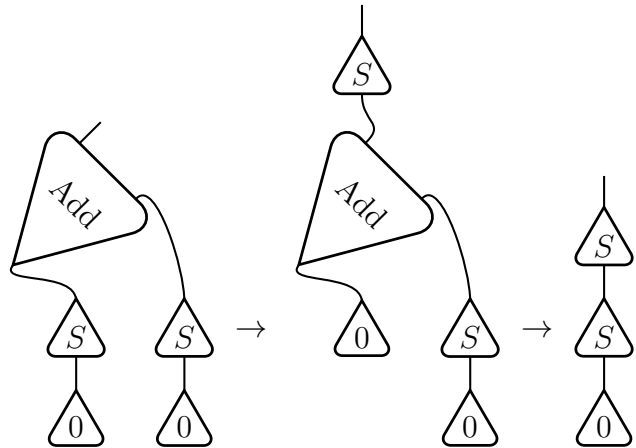
\begin{tikzpicture}[baseline=(0)]
\inetcell(0){$0$}[R]
\inetwirefree(0.pal)
\end{tikzpicture}
\qquad
\begin{tikzpicture}[baseline=(S)]
\inetcell(S){$S$}[R]
\inetwirefree(S.pal)
\inetwirefree(S.middle pax)
\end{tikzpicture}
\qquad
\begin{tikzpicture}[baseline=(Add)]
\inetcell(Add){$\text{Add}$}[R]
\inetwirefree(Add.pal)
\inetwirefree(Add.left pax)
\inetwirefree(Add.right pax)
\end{tikzpicture}
$$
Теперь определим правила взаимодействия.
В нашем случае, мы можем тривиальным образом отобразить определение $\text{Add}$.
$$
\begin{tikzpicture}[baseline=(p.base)]
\matrix[column sep=1em]{
\inetcell(0){$0$}[R] &
\inetcell(Add){$\text{Add}$}[L] \\ };
\node (p) [right=of 0.pal] {$\phantom x$};
\inetwirefree(Add.left pax)
\inetwirefree(Add.right pax)
\inetwire(0.pal)(Add.pal)
\end{tikzpicture}
\rightarrow
\begin{tikzpicture}[baseline=(p.base)]
\matrix[row sep=1em]{
\node (t) {$\phantom x$}; \\
\node (p) {$\phantom x$}; \\
\node (b) {$\phantom x$}; \\ };
\inetwirecoords(t)(b)
\end{tikzpicture}
\qquad
\begin{tikzpicture}[baseline=(p.base)]
\matrix[column sep=1em]{
\inetcell(S){$S$}[R] &
\inetcell(Add){$\text{Add}$}[L] \\ };
\node (p) [right=of S.pal] {$\phantom x$};
\inetwirefree(S.middle pax)
\inetwirefree(Add.left pax)
\inetwirefree(Add.right pax)
\inetwire(0.pal)(Add.pal)
\end{tikzpicture}
\rightarrow
\begin{tikzpicture}[baseline=(p.base)]
\inetcell(Add){$\text{Add}$}[L]
\inetcell[node distance=1em, right=of Add.right pax](S){$S$}[R]
\node (p) [right=of Add.middle pax] {$\phantom x$};
\inetwirefree(Add.pal)
\inetwirefree(Add.left pax)
\inetwirefree(S.pal)
\inetwire(Add.right pax)(S.middle pax)
\end{tikzpicture}
$$
Обращаем внимание на сохранение интерфейса при взаимодействии.

Теперь рассмотрим сеть, соответствующую выражению $\text{Add}(S(0), S(0))$.
$$
\begin{tikzpicture}
\matrix[row sep=1em]{
\inetcell[rotate=-45](Add){$\text{Add}$}[D] & \\
\inetcell(S1){$S$}[U] & \inetcell(S2){$S$}[U] \\
\inetcell(01){$0$}[U] & \inetcell(02){$0$}[U] \\ };
\inetwirefree(Add.right pax)
\inetwire(S1.pal)(Add.pal)
\inetwire(S2.pal)(Add.left pax)
\inetwire(01.pal)(S1.middle pax)
\inetwire(02.pal)(S2.middle pax)
\end{tikzpicture}
$$
В этом примере возможна лишь одна последовательность редукций, так как на каждом шаге имеется лишь одна активная пара.
Полная последовательность редукций показана ниже.
Результатом будет сеть, представляющая выражение $S(S(0))$.
$$
\begin{tikzpicture}[baseline=(S2)]
\matrix[row sep=1em]{
\inetcell[rotate=-45](Add){$\text{Add}$}[D] & \\
\inetcell(S1){$S$}[U] & \inetcell(S2){$S$}[U] \\
\inetcell(01){$0$}[U] & \inetcell(02){$0$}[U] \\ };
\inetwirefree(Add.right pax)
\inetwire(S1.pal)(Add.pal)
\inetwire(S2.pal)(Add.left pax)
\inetwire(01.pal)(S1.middle pax)
\inetwire(02.pal)(S2.middle pax)
\end{tikzpicture}
\rightarrow
\begin{tikzpicture}[baseline=(S2)]
\matrix[row sep=1em]{
\inetcell(S1){$S$}[U] & \\
\inetcell[rotate=-45](Add){$\text{Add}$}[D] & \\
\inetcell(01){$0$}[U] & \inetcell(S2){$S$}[U] \\
& \inetcell(02){$0$}[U] \\ };
\inetwirefree(S1.pal)
\inetwire(01.pal)(Add.pal)
\inetwire(S2.pal)(Add.left pax)
\inetwire(S1.middle pax)(Add.right pax)
\inetwire(02.pal)(S2.middle pax)
\end{tikzpicture}
\rightarrow
\begin{tikzpicture}[baseline=(S2)]
\matrix[row sep=1em]{
\inetcell(S1){$S$}[U] \\
\inetcell(S2){$S$}[U] \\
\inetcell(02){$0$}[U] \\ };
\inetwirefree(S1.pal)
\inetwire(S2.pal)(S1.middle pax)
\inetwire(02.pal)(S2.middle pax)
\end{tikzpicture}
$$

Предыдущий пример слишком прост, чтобы показать существенные особенности сетей взаимодействия.
Более интересной будет система взаимодействия, реализующая операцию умножения чисел:
\begin{align*}
\text{Mult}(0, y) &= 0; \\
\text{Mult}(S(x), y) &= \text{Add}(\text{Mult}(x, y), y).
\end{align*}
Чтобы представить этот алгоритм, нам потребуется новый агент $\text{Mult}$ перемножения чисел, правила взаимодействия для которого сложнее, чем у агента $\text{Add}$, так как умножение не является линейной операцией (как в случае со сложением).
Для того чтобы оставаться в соответствии с определением правил взаимодействия, мы должны сохранить интерфейс.
Следующая схема иллюстрирует правила для $\text{Mult}$.
$$
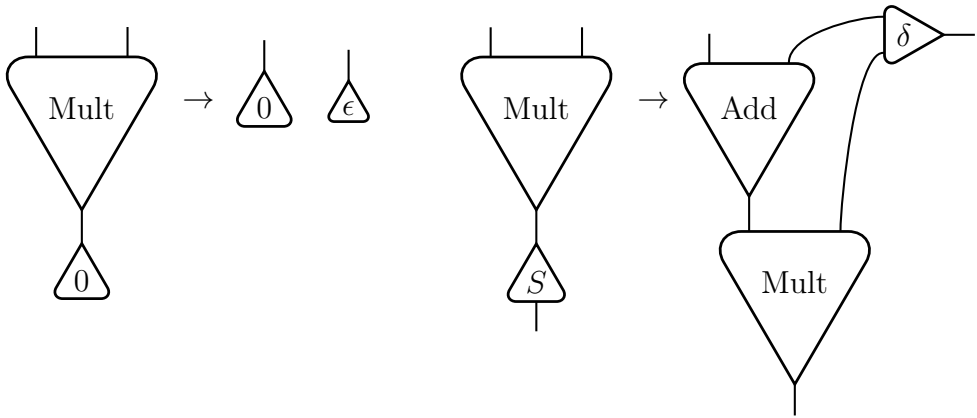
\begin{tikzpicture}[baseline=(Mult)]
\matrix[row sep=1em]{
\inetcell(Mult){$\text{Mult}$}[D] \\
\inetcell(0){$0$}[U] \\ };
\inetwirefree(Mult.left pax)
\inetwirefree(Mult.right pax)
\inetwire(0.pal)(Mult.pal)
\end{tikzpicture}
\rightarrow
\begin{tikzpicture}[baseline=(0)]
\matrix[column sep=1em]{
\inetcell(0){$0$}[U] &
\inetcell(e){$\epsilon$}[U] \\ };
\inetwirefree(0.pal)
\inetwirefree(e.pal)
\end{tikzpicture}
\qquad
\begin{tikzpicture}[baseline=(Mult)]
\matrix[row sep=1em]{
\inetcell(Mult){$\text{Mult}$}[D] \\
\inetcell(S){$S$}[U] \\ };
\inetwirefree(Mult.left pax)
\inetwirefree(Mult.right pax)
\inetwirefree(S.middle pax)
\inetwire(S.pal)(Mult.pal)
\end{tikzpicture}
\rightarrow
\begin{tikzpicture}[baseline=(Add)]
\inetcell(Mult){$\text{Mult}$}[D]
\inetcell[node distance=1em, above=of Mult.right pax](Add){$\text{Add}$}[D]
\inetcell[node distance=3em, right=of Add.above left pax](d){$\delta$}[R]
\inetwirefree(Mult.pal)
\inetwirefree(Add.right pax)
\inetwirefree(d.pal)
\inetwire(Mult.right pax)(Add.pal)
\inetwire(Add.left pax)(d.left pax)
\inetwire(Mult.left pax)(d.right pax)
\end{tikzpicture}
$$
Чтобы сохранить интерфейс, мы воспользовались удаляющим $\epsilon$ и дублирующим $\delta$ агентами, которые были введены в примере~\ref{erasedup}.

Последний пример демонстрирует один из самых важных аспектов сетей взаимодействия.
Заметим, что дублирование активных пар невозможно, поэтому взаимодействие активных пар происходит лишь однажды; иначе вычисление по сути производилось бы дважды и, следовательно, было бы неоптимальным.
Действительно, чтобы продублировать сеть, $\delta$ должен провзаимодействовать со всеми агентами в сети, но если $\alpha$ и $\beta$ соединены своими главными портами, то они не могут взаимодействовать с $\delta$ и, стало быть, не могут быть продублированы.

Приведем еще один пример арифметической операции.

\begin{example}
\label{maxexample}
Рассмотрим функцию, возвращающую наибольшее из двух чисел:
\begin{align*}
\text{Max}(0, y) &= y; \\
\text{Max}(x, 0) &= x; \\
\text{Max}(S(x), S(y)) &= S(\text{Max}(x, y)).
\end{align*}
С данным определением связана одна проблема: в нем используются одновременно оба аргумента функции $\text{Max}$.
Дело в том, что если мы попытаемся действовать аналогично двум предыдущим примерам операций на числах, нам потребуется два главных порта для агента $\text{Max}$, но это невозможно в сетях взаимодействия.
Тем не менее, мы можем изменить определение $\text{Max}$, введя новую функцию $\text{Max}'$, чтобы получить эквивалентную систему, где каждая операция определена с использованием только одного аргумента:
\begin{align*}
\text{Max}(0, y) &= y; \\
\text{Max}(S(x), y) &= \text{Max}'(x, y); \\
\text{Max}'(x, 0) &= S(x); \\
\text{Max}'(x, S(y)) &= S(\text{Max}(x, y)). \\
\end{align*}
Начнем с правил взаимодействия для агента $\text{Max}$.
$$
\begin{tikzpicture}[baseline=(p.base)]
\matrix[column sep=1em]{
\inetcell(0){$0$}[R] &
\inetcell(Max){$\text{Max}$}[L] \\ };
\node (p) [right=of 0.pal] {$\phantom x$};
\inetwirefree(Max.left pax)
\inetwirefree(Max.right pax)
\inetwire(0.pal)(Max.pal)
\end{tikzpicture}
\rightarrow
\begin{tikzpicture}[baseline=(p.base)]
\matrix[row sep=1em]{
\node (t) {$\phantom x$}; \\
\node (p) {$\phantom x$}; \\
\node (b) {$\phantom x$}; \\ };
\inetwirecoords(t)(b)
\end{tikzpicture}
\qquad
\begin{tikzpicture}[baseline=(p.base)]
\matrix[column sep=1em]{
\inetcell(S){$S$}[R] &
\inetcell(Max){$\text{Max}$}[L] \\ };
\node (p) [right=of S.pal] {$\phantom x$};
\inetwirefree(Max.left pax)
\inetwirefree(Max.right pax)
\inetwirefree(S.middle pax)
\inetwire(S.pal)(Max.pal)
\end{tikzpicture}
\rightarrow
\begin{tikzpicture}[baseline=(Max)]
\inetcell[rotate=-45](Max){$\text{Max}'$}[R]
\inetcell[draw=none, node distance=2em, above right=of Max.left pax](t){$\phantom S$}[R]
\inetwirefree(Max.pal)
\inetwirefree(Max.right pax)
\inetwire(Max.left pax)(t.middle pax)
\end{tikzpicture}
$$
Теперь нам остается определить правила для $\text{Max}'$.
$$
\begin{tikzpicture}[baseline=(p.base)]
\matrix[column sep=1em]{
\inetcell(Max){$\text{Max}'$}[R] &
\inetcell(0){$0$}[L] \\ };
\node (p) [right=of 0.pal] {$\phantom x$};
\inetwirefree(Max.left pax)
\inetwirefree(Max.right pax)
\inetwire(0.pal)(Max.pal)
\end{tikzpicture}
\rightarrow
\begin{tikzpicture}[baseline=(S)]
\inetcell(S){$S$}[U]
\inetwirefree(S.pal)
\inetwirefree(S.middle pax)
\end{tikzpicture}
\qquad
\begin{tikzpicture}[baseline=(p.base)]
\matrix[column sep=1em]{
\inetcell(Max){$\text{Max}'$}[R] &
\inetcell(S){$S$}[L] \\ };
\node (p) [right=of S.middle pax] {$\phantom x$};
\inetwirefree(Max.left pax)
\inetwirefree(Max.right pax)
\inetwirefree(S.middle pax)
\inetwire(S.pal)(Max.pal)
\end{tikzpicture}
\rightarrow
\begin{tikzpicture}[baseline=(p.base)]
\inetcell[rotate=45](Max){$\text{Max}$}[L]
\inetcell[node distance=2em, above left=of Max.right pax](S){$S$}[L]
\node (p) [left=of S.pal] {$\phantom x$};
\inetwirefree(S.pal)
\inetwirefree(Max.pal)
\inetwirefree(Max.left pax)
\inetwire(S.middle pax)(Max.right pax)
\end{tikzpicture}
$$
Построение же системы взаимодействия, способной вычислять наименьшее из двух натуральных чисел, мы оставим читателю (см.~параграф~\ref{exercise}).
\end{example}

Пример~\ref{maxexample} дает нам способ представлять произвольные арифметические операции.
На самом деле, в следующем параграфе мы увидим, что любые <<чистые>> функции можно компилировать в сети взаимодействия, которые сами таким образом оказываются своеобразным универсальным языком программирования.

\section{Полнота по Тьюрингу}
\label{complete}

Модель вычислений полна по Тьюрингу, если любая вычислимая функция представима в ней.
В случае сетей взаимодействия, доказать полноту по Тьюрингу можно, например, представив комбинаторную логику~---~систему комбинаторов с константами $S$ и $K$ и двумя правилами редукции, имеющую ту же мощность с точки зрения вычислений, что и $\lambda$-исчисление:
\begin{align*}
K\ x\ y &\rightarrow x; \\
S\ x\ y\ z &\rightarrow x\ z\ (y\ z).
\end{align*}
Чтобы представить эту систему в сетях взаимодействия, нам потребуется агент $@$, соответствующий аппликации, а также несколько агентов, представляющих сами комбинаторы.
В частности, комбинатор $K$ будет представлен двумя агентами $K_0$ и $K_1$ со следующими правилами взаимодействия.
$$
\begin{tikzpicture}[baseline=(p.base)]
\matrix[column sep=1em]{
\inetcell(k){$K_0$}[R] &
\inetcell(a){$@$}[L] \\ };
\node (p) [right=of a.pal] {$\phantom x$};
\inetwirefree(a.left pax)
\inetwirefree(a.right pax)
\inetwire(a.pal)(k.pal)
\end{tikzpicture}
\rightarrow
\begin{tikzpicture}[baseline=(k)]
\inetcell(k){$K_1$}[U]
\inetwirefree(k.pal)
\inetwirefree(k.middle pax)
\end{tikzpicture}
\qquad
\begin{tikzpicture}[baseline=(p.base)]
\matrix{
\inetcell(a){$@$}[D] \\
\node (p) {$\phantom x$}; \\
\inetcell(k){$K_1$}[U] \\ };
\inetwirefree(a.left pax)
\inetwirefree(a.right pax)
\inetwirefree(k.middle pax)
\inetwire(a.pal)(k.pal)
\end{tikzpicture}
\rightarrow
\begin{tikzpicture}[baseline=(p.base)]
\matrix[row sep=2em]{
\inetcell[opacity=0](k){$\phantom{K_1}$}[U] &
\inetcell(e){$\epsilon$}[U] \\ };
\node (p) [left=of e] {$\phantom x$};
\inetwirefree(e.pal)
\inetwirecoords(k.above middle pax)(k.above pal)
\end{tikzpicture}
$$
Комбинатор $S$ может быть представлен аналогичным образом тремя агентами и тремя правилами; мы оставим эту задачу читателю в качестве упражнения.

Сети взаимодействия используются и для реализации $\lambda$-исчисления.
Действительно, первая реализация оптимальной стратегии для $\lambda$-исчисления (т.~е. стратегии, которая выполняет наименьшее число $\beta$-редукций, необходимое для достижения нормальной формы терма) формально являлась системой взаимодействия.
Сети взаимодействия также используются и для других (неоптимальных, но в некоторых случаях более эффективных) реализациях $\lambda$-исчисления.

На самом деле, если мы ограничимся линейным $\lambda$-исчислением, то нам потребуются только два агента $@$ и $\lambda$, представляющих, аппликацию и абстракцию, соответственно; при этом роль переменных играют связи.
В этом случае $\beta$-редукция будет соответствовать следующему правилу взаимодействия.
$$
\begin{tikzpicture}[baseline=(p.base)]
\matrix{
\inetcell(a){$@$}[D] \\
\node (p) {$\phantom x$}; \\
\inetcell(l){$\lambda$}[U] \\ };
\inetwirefree(a.left pax)
\inetwirefree(a.right pax)
\inetwirefree(l.left pax)
\inetwirefree(l.right pax)
\inetwire(a.pal)(l.pal)
\end{tikzpicture}
\rightarrow
\begin{tikzpicture}[baseline=(p.base)]
\matrix{
\inetcell[opacity=0](a){$\phantom @$}[D] \\
\node (p) {$\phantom x$}; \\
\inetcell[opacity=0](l){$\phantom \lambda$}[U] \\ };
\inetwirecoords(a.above left pax)(l.above left pax)
\inetwirecoords(a.above right pax)(l.above right pax)
\end{tikzpicture}
$$

Чтобы представить произвольные $\lambda$-термы, нам пришлось бы ввести агенты для копирования и удаления.
Также нам понадобились бы дополнительные агенты, которые бы могли контролировать область действия абстракций.

\section{Представление списков}

Упорядоченные последовательности элементов, то есть списки, можно представить разными способами.
В частности, мы можем построить список с помощью бинарного агента $\text{Cons}$, который будет соединять первый элемент списка с остальной его частью.
Пустой же список может быть представлен с помощью агента $\text{Nil}$.
Данный способ соответствует традиционному определению этой структуры данных в функциональных языках программирования.
Соединение двух списков, то есть операцию конкатенации, определяют следующим образом:
\begin{align*}
\text{Append}(\text{Nil}, l) &= l; \\
\text{Append}(\text{Cons}(x, l), l') &= \text{Cons}(x, \text{Append}(l, l')).
\end{align*}
Для конкатенации двух списков, определенных таким образом, требуется время $O(n)$, пропорциональное длине $n$ первого из списков.

Читателю не составит труда представить определение выше в сетях взаимодействия с помощью трех агентов $\text{Nil}$, $\text{Cons}$ и $\text{Append}$.
Однако, если мы примем во внимание тот факт, что сети взаимодействия представляют собой графы и не обязаны быть деревьями, то получим более эффективную реализацию.
Идея заключается в том, чтобы иметь связи как с началом списка, так и с его концом, то есть в использовании так называемых разностных списков.
Это можно сделать с помощью бинарного агента $\text{Diff}$, который бы имел связи с двумя агентами $\text{Cons}$, соответствующими первому и последнему элементам списка.

При этом пустой список $\text{Nil}$ будет представляться следующей сетью.
$$
\begin{tikzpicture}
\inetcell(d){$\text{Diff}$}[U]
\inetwirefree(d.pal)
\inetwire(d.left pax)(d.right pax)
\end{tikzpicture}
$$

Итак, представим эффективную версию конкатенации списков с помощью двух правил взаимодействия.
Заметим, что конкатенация двух списков будет выполняться за фиксированное время $O(1)$ вне зависимости от их длины.
$$
\begin{tikzpicture}[baseline=(p.base)]
\matrix{
\inetcell(a){$\text{Append}$}[D] \\
\node (p) {$\phantom x$}; \\
\inetcell(l){$\text{Diff}$}[U] \\ };
\inetwirefree(a.left pax)
\inetwirefree(a.right pax)
\inetwirefree(l.left pax)
\inetwirefree(l.right pax)
\inetwire(a.pal)(l.pal)
\end{tikzpicture}
\rightarrow
\begin{tikzpicture}[baseline=(p.base)]
\matrix{
\inetcell(a){$\text{Diff}$}[U] &
\node (p) {$\phantom x$}; &
\inetcell(l){$\text{Open}$}[U] \\ };
\inetwirefree(a.left pax)
\inetwirefree(l.right pax)
\inetwirefree(a.pal)
\inetwirefree(l.pal)
\inetwire(a.right pax)(l.left pax)
\end{tikzpicture}
\qquad
\begin{tikzpicture}[baseline=(p.base)]
\matrix{
\inetcell(a){$\text{Open}$}[D] \\
\node (p) {$\phantom x$}; \\
\inetcell(l){$\text{Diff}$}[U] \\ };
\inetwirefree(a.left pax)
\inetwirefree(a.right pax)
\inetwirefree(l.left pax)
\inetwirefree(l.right pax)
\inetwire(a.pal)(l.pal)
\end{tikzpicture}
\rightarrow
\begin{tikzpicture}[baseline=(p.base)]
\matrix{
\inetcell[opacity=0](a){$\text{Open}$}[D] \\
\node (p) {$\phantom x$}; \\
\inetcell[opacity=0](l){$\text{Diff}$}[U] \\ };
\inetwirecoords(a.above left pax)(l.above right pax)
\inetwirecoords(a.above right pax)(l.above left pax)
\end{tikzpicture}
$$

Проиллюстрируем полученную систему взаимодействия примером конкатенации двух списков: вычисление будет состоять всего из двух редукций.
$$
\begin{tikzpicture}[baseline=(d)]
\matrix[row sep=1em]{
& & & \inetcell[rotate=-45](a){$\text{Append}$}[D] & \\
& & \inetcell(d){$\text{Diff}$}[U] & &
\inetcell(l){$\text{Diff}$}[U] \\
\inetcell(1){$1$}[D] &
\inetcell(c1){$\text{Cons}$}[U] & & &
\inetcell(c3){$\text{Cons}$}[L] \\
\inetcell(2){$2$}[D] &
\inetcell(c2){$\text{Cons}$}[U] &
\inetcell[opacity=0](p){$\text{Diff}$}[U] & &
\inetcell(3){$3$}[U] \\ };
\inetwirefree(a.right pax)
\inetwire(l.pal)(a.left pax)
\inetwire(d.pal)(a.pal)
\inetwirecoords(d.right pax)(p.right pax)
\inetwire(c1.pal)(d.left pax)
\inetwire(c2.pal)(c1.right pax)
\inetwire(c3.pal)(l.left pax)
\inetwire(1.pal)(c1.left pax)
\inetwire(2.pal)(c2.left pax)
\inetwire(3.pal)(c3.left pax)
\inetwire(p.right pax)(c2.right pax)
\inetwire(l.right pax)(c3.right pax)
\end{tikzpicture}
\rightarrow
$$
$$
\begin{tikzpicture}[baseline=(d)]
\matrix[row sep=1em]{
& \inetcell(d){$\text{Diff}$}[U] &
\inetcell(o){$\text{Open}$}[U] &
\inetcell(l){$\text{Diff}$}[U] \\
\inetcell(1){$1$}[D] &
\inetcell(c1){$\text{Cons}$}[U] & &
\inetcell(c3){$\text{Cons}$}[L] \\
\inetcell(2){$2$}[D] &
\inetcell(c2){$\text{Cons}$}[U] &
\inetcell[opacity=0](p){$\text{Open}$}[U] &
\inetcell(3){$3$}[U] \\ };
\inetwirefree(d.pal)
\inetwire(l.pal)(o.pal)
\inetwire(d.right pax)(o.left pax)
\inetwirecoords(o.right pax)(p.right pax)
\inetwire(c1.pal)(d.left pax)
\inetwire(c2.pal)(c1.right pax)
\inetwire(c3.pal)(l.left pax)
\inetwire(1.pal)(c1.left pax)
\inetwire(2.pal)(c2.left pax)
\inetwire(3.pal)(c3.left pax)
\inetwire(p.right pax)(c2.right pax)
\inetwire(l.right pax)(c3.right pax)
\end{tikzpicture}
\rightarrow
\begin{tikzpicture}[baseline=(d)]
\matrix[row sep=1em]{
& & \inetcell(d){$\text{Diff}$}[U] \\
\inetcell(1){$1$}[D] & \inetcell(c1){$\text{Cons}$}[U] & \\
\inetcell(2){$2$}[D] & \inetcell(c2){$\text{Cons}$}[U] & \\
\inetcell(3){$3$}[D] & \inetcell(c3){$\text{Cons}$}[U] & \inetcell[opacity=0](p){$\text{Diff}$}[U] \\ };
\inetwirefree(d.pal)
\inetwirecoords(d.right pax)(p.right pax)
\inetwire(c1.pal)(d.left pax)
\inetwire(c2.pal)(c1.right pax)
\inetwire(c3.pal)(c2.right pax)
\inetwire(1.pal)(c1.left pax)
\inetwire(2.pal)(c2.left pax)
\inetwire(3.pal)(c3.left pax)
\inetwire(p.right pax)(c3.right pax)
\end{tikzpicture}
$$

\section{Комбинаторы взаимодействия}

Подобно тому, как $K$ и $S$ в комбинаторной логике способны представить любые вычислимые функции, агенты $\delta$, $\gamma$ и $\epsilon$ порождают универсальную систему взаимодействия.
Эти три агента показаны на рисунке~\ref{inetcomb}.
Первые два из них осуществляют мультиплексирование (т.~е. объединение двух связей в одну), а третий~---~удаление.
Оказывается, соединяя определенным образом исключительно эти три комбинатора взаимодействия, можно заменить вообще любую сеть взаимодействия.

\begin{figure}[h]
$$
\begin{tikzpicture}[baseline=(c)]
\inetcell(c){$\gamma$}
\inetwirefree(c.pal)
\inetwirefree(c.left pax)
\inetwirefree(c.right pax)
\end{tikzpicture}
\qquad
\begin{tikzpicture}[baseline=(c)]
\inetcell(c){$\delta$}
\inetwirefree(c.pal)
\inetwirefree(c.left pax)
\inetwirefree(c.right pax)
\end{tikzpicture}
\qquad
\begin{tikzpicture}[baseline=(c)]
\inetcell(c){$\epsilon$}
\inetwirefree(c.pal)
\end{tikzpicture}
$$
\caption{комбинаторы взаимодействия ($\gamma$, $\delta$ и $\epsilon$).}
\label{inetcomb}
\end{figure}

Система комбинаторов имеет шесть правил взаимодействия, которые показаны на рисунке~\ref{combrule}.
Очевидно, что агент $\epsilon$ ведет себя как операция удаления, <<съедая>> все, с чем он взаимодействует.
В свою очередь, мультиплексирующие агенты либо аннигилируют (если агенты совпадают), либо дублируют друг друга (если они различны).
Заметим, что в последнем правиле взаимодействия правая часть является пустой сетью.

Система комбинаторов взаимодействия \textit{универсальна} в том смысле, что внутри нее можно представить любую другую систему взаимодействия.
Впрочем, известны также и другие универсальные системы взаимодействия, хотя они так или иначе сложнее системы комбинаторов.

\begin{figure}[h]
$$
\begin{tikzpicture}[baseline=(p.base)]
\matrix{
\inetcell(t){$\delta$}[D] \\
\node (p) {$\phantom I$}; \\
\inetcell(b){$\delta$}[U] \\ };
\inetwirefree(t.left pax)
\inetwirefree(t.right pax)
\inetwirefree(b.left pax)
\inetwirefree(b.right pax)
\inetwire(t.pal)(b.pal)
\end{tikzpicture}
\rightarrow
\begin{tikzpicture}[baseline=(p.base)]
\matrix{
\inetcell[opacity=0](t){$\delta$}[D] \\
\node (p) {$\phantom I$}; \\
\inetcell[opacity=0](b){$\delta$}[U] \\ };
\inetwirecoords(t.above left pax)(b.above left pax)
\inetwirecoords(t.above right pax)(b.above right pax)
\end{tikzpicture}
\qquad
\begin{tikzpicture}[baseline=(p.base)]
\matrix{
\inetcell(t){$\gamma$}[D] \\
\node (p) {$\phantom I$}; \\
\inetcell(b){$\gamma$}[U] \\ };
\inetwirefree(t.left pax)
\inetwirefree(t.right pax)
\inetwirefree(b.left pax)
\inetwirefree(b.right pax)
\inetwire(t.pal)(b.pal)
\end{tikzpicture}
\rightarrow
\begin{tikzpicture}[baseline=(p.base)]
\matrix{
\inetcell[opacity=0](t){$\gamma$}[D] \\
\node (p) {$\phantom I$}; \\
\inetcell[opacity=0](b){$\gamma$}[U] \\ };
\inetwirecoords(t.above left pax)(b.above right pax)
\inetwirecoords(t.above right pax)(b.above left pax)
\end{tikzpicture}
$$
$$
\begin{tikzpicture}[baseline=(p.base)]
\matrix{
\inetcell(t){$\gamma$}[D] \\
\node (p) {$\phantom I$}; \\
\inetcell(b){$\delta$}[U] \\ };
\inetwirefree(t.left pax)
\inetwirefree(t.right pax)
\inetwirefree(b.left pax)
\inetwirefree(b.right pax)
\inetwire(t.pal)(b.pal)
\end{tikzpicture}
\rightarrow
\begin{tikzpicture}[baseline=(p.base)]
\matrix{
\inetcell(tl){$\delta$}[U] & & \inetcell(tr){$\delta$}[U] \\
& \node (p) {$\phantom I$}; & \\
\inetcell(bl){$\gamma$}[D] & & \inetcell(br){$\gamma$}[D] \\ };
\inetwirefree(tl.pal)
\inetwirefree(tr.pal)
\inetwirefree(bl.pal)
\inetwirefree(br.pal)
\inetwire(tl.left pax)(bl.right pax)
\inetwire(tl.right pax)(br.right pax)
\inetwire(tr.left pax)(bl.left pax)
\inetwire(tr.right pax)(br.left pax)
\end{tikzpicture}
$$
$$
\begin{tikzpicture}[baseline=(p.base)]
\matrix{
\inetcell(t){$\delta$}[D] \\
\node (p) {$\phantom I$}; \\
\inetcell(b){$\epsilon$}[U] \\ };
\inetwirefree(t.left pax)
\inetwirefree(t.right pax)
\inetwire(t.pal)(b.pal)
\end{tikzpicture}
\rightarrow
\begin{tikzpicture}[baseline=(p.base)]
\matrix{
\inetcell(t){$\epsilon$}[U] &
\node (p) {$\phantom I$}; &
\inetcell(b){$\epsilon$}[U] \\ };
\inetwirefree(t.pal)
\inetwirefree(b.pal)
\end{tikzpicture}
\qquad
\begin{tikzpicture}[baseline=(p.base)]
\matrix{
\inetcell(t){$\gamma$}[D] \\
\node (p) {$\phantom I$}; \\
\inetcell(b){$\epsilon$}[U] \\ };
\inetwirefree(t.left pax)
\inetwirefree(t.right pax)
\inetwire(t.pal)(b.pal)
\end{tikzpicture}
\rightarrow
\begin{tikzpicture}[baseline=(p.base)]
\matrix{
\inetcell(t){$\epsilon$}[U] &
\node (p) {$\phantom I$}; &
\inetcell(b){$\epsilon$}[U] \\ };
\inetwirefree(t.pal)
\inetwirefree(b.pal)
\end{tikzpicture}
\qquad
\begin{tikzpicture}[baseline=(p.base)]
\matrix{
\inetcell(t){$\epsilon$}[D] \\
\node (p) {$\phantom I$}; \\
\inetcell(b){$\epsilon$}[U] \\ };
\inetwire(t.pal)(b.pal)
\end{tikzpicture}
\rightarrow
\begin{tikzpicture}[baseline=(p.base)]
\matrix{
\inetcell[opacity=0](t){$\epsilon$}[U] &
\node (p) {$\phantom I$}; &
\inetcell[opacity=0](b){$\epsilon$}[U] \\ };
\end{tikzpicture}
$$
\caption{правила взаимодействия для комбинаторов.}
\label{combrule}
\end{figure}

\section{Текстовая запись и стратегии}

Для сетей взаимодействия вполне естественно графическое представление, и зачастую схемы гораздо легче понять, чем их текстовое определение.
Однако, формальная текстовая запись имеет значительные преимущества: она упрощает написание программ, так как графические редакторы не всегда применимы, а также дает возможность задавать свойства сетей более кратко и элегантно.
Были разработаны уже несколько форм текстовой записи для сетей взаимодействия.
Ниже мы пройдем путь от примитивных, но избыточных обозначений к более компактной, но нетривиальной записи, а затем введем на основе последней исчисление взаимодействия.

В качестве примера мы будем использовать одну и ту же сеть, показанную на рисунке~\ref{lterm}, применяя правило взаимодействия из параграфа~\ref{complete}, которое соответствует $\beta$-редукции в линейном $\lambda$-исчислении.

\begin{figure}[b]
$$
\begin{tikzpicture}
\matrix[column sep=1em]{
\inetcell(a){$@$}[D] &
\inetcell(p){$\lambda$}[U] &
\inetcell(f){$\lambda$}[D] \\ };
\inetwirefree(a.right pax)
\inetwire(a.pal)(f.pal)
\inetwire(a.left pax)(p.pal)
\inetwire(p.left pax)(p.right pax)
\inetwire(f.left pax)(f.right pax)
\end{tikzpicture}
$$
\caption{$\lambda$-терм, представленный в виде сети.}
\label{lterm}
\end{figure}

Естественные обозначения можно получить простым перечислением всех агентов, участвующих в сети, используя определенные соглашения.
Например, мы могли бы договориться всегда перечислять порты агентов против часовой стрелки, начиная с главного порта.
Связь можно задать, используя одно и то же имя для двух портов.
Тогда сеть, представленная на рисунке~\ref{lterm}, будет выглядеть следующим образом:
$$
@(a, b, c), \lambda(a, d, d), \lambda(b, e, e).
$$
Действительно, здесь два агента $\lambda$, представляющих абстракцию, и один агент $@$, представляющий аппликацию.
Заметим, что вхождение одного и того же имени для двух разных портов обозначает связь между ними; например, в записи $\lambda(a, e, e)$ вхождение $e$ дважды означает, что в сети имеется связь между двумя дополнительными портами данного агента $\lambda$.

Те же обозначения можно использовать и для записи правил взаимодействия.
К примеру, правило взаимодействия для $\beta$-редукции в линейном $\lambda$-исчислении мы могли бы записать как
$$
@(a, b, c), \lambda(a, d, e) \rightarrow I(b, d), I(c, e),
$$
где символ $I$ используется для представления связей с интерфейсом сети (а не агентами).
Следует обратить внимание, что в левой и правой частях правила используются одни и те же имена для портов и связей, так как правила взаимодействия по определению обязаны сохранять интерфейс сети.

Еще один способ записи заключается в том, чтобы использовать порядковые номера портов вместо имен, начиная с $0$ для главного порта.
Произвольную сеть можно преставить в виде упорядоченной пары $(A, W)$, где $A$~---~множество входящих в сеть агентов, а $W$~---~множество связей.

Например, агенты, из которых состоит сеть на рисунке~\ref{lterm}, составляют множество
$A = \{@_1, \lambda_1, \lambda_2\}$, поэтому сеть записывается следующим образом:
$$
(\{@_1, \lambda_1, \lambda_2\}, \{@_1.0 = \lambda_1.0, @_1.1 = \lambda_2.0, \lambda_1.1 = \lambda_1.2, \lambda_2.1 = \lambda_2.2\})
$$

В свою очередь, правило взаимодействия для $\beta$-редукции будет иметь вид
$$
(\{\lambda_i, @_j\}, \{\lambda_i.0 = @_j.0\}) \rightarrow (\varnothing, \{\lambda_i.1 = @_j.1, \lambda.2 = @_j.2\}).
$$

Наконец, третий способ записи, который оказывается более кратким, основан на представлении активных пар в виде уравнений.
В нашем случае пример имеет форму
$$
\lambda(a, a) = @(\lambda(b, b), c),
$$
где знак равенства означает связь между главным портом агента $\lambda$ в левой части уравнения и главным портом агента $@$~---~в правой его части (т.~е. уравнение представляет активную пару $\lambda \bowtie @$).
Левая часть уравнения $\lambda(a, a)$ означает, что оба дополнительных порта этого агента связаны, ровно как и в случае $\lambda(b, b)$.

Воспользуемся аналогичным обозначением для правил взаимодействия, заменив знак равенства на <<$\bowtie$>>, а круглые скобки~---~квадратными, чтобы отличать активную пару от результата взаимодействия.
Тогда правило $\beta$-редукции примет вид
$$
@[x, y] \bowtie \lambda[x, y].
$$

Будучи чрезвычайно сжатой формой записи, введенные обозначения активных пар и правил взаимодействия оказываются более удобными при реализации систем взаимодействия.
Они также послужат основой для исчисления взаимодействия, к построению которого мы сейчас перейдем.

\subsection{Исчисление взаимодействия}

Ранее уже говорилось, что системы взаимодействия обладают свойством сильной конфлюэнтности, но, как и в других системах, выполняющих редукцию, для систем взаимодействия существуют различные определения нормальных форм и стратегий (например, когда говорят о так называемых <<ленивых>> вычислениях, часто имеют в виду слабую нормальную форму).
Ниже мы увидим, что исчисление взаимодействия дает возможность точно определять такие понятия.

Сети взаимодействия оказали наибольшее влияние на $\lambda$-исчисление, в котором стратегии играют важную роль: только нормализующие стратегии там гарантируют получение нормальной формы произвольного $\lambda$-терма, если таковая существует.

Начнем с описания синтаксиса исчисления взаимодействия.

\begin{description}
\item[Агенты.] Пусть $\Sigma$~---~множество символов $\alpha, \beta, \dots$ и каждый символ имеет \textit{арность} (мы предполагаем наличие функции $\text{Ar}: \Sigma \rightarrow \mathbb N$, возвращающей арность).
Вхождение символа будем называть \textit{агентом}.
Арность символа соответствует числу дополнительных портов соответствующего агента.

\item[Имена.] Пусть $N$~---~множество имен $x, y, z, \dots$, которое не пересекается с $\Sigma$.

\item[Термы.] Термы строятся с помощью агентов из $\Sigma$ и имен из $N$ согласно грамматике
$$
t ::= x\ |\ \alpha(t_1, \dots, t_n),
$$
где $x \in N$, $\alpha \in \Sigma$ и $\text{Ar}(\alpha) = n$, причем каждое имя может иметь не более двух вхождений в терм.
Если $n = 0$, то скобки опускаются.

Если имя имеет два вхождения, говорят, что оно \textit{связано}; в противном случае имя \textit{свободно}.
Свободные имена могут иметь лишь одно вхождение, поэтому термы \textit{линейны} в том же смысле, что и термы в линейном $\lambda$-исчислении.

Последовательность термов $t_1, \dots, t_n$ мы будем сокращенно обозначать $\vec t$.

Терм вида $\alpha(\vec t)$ можно представить в форме дерева: главный порт $\alpha$ служит корнем, а термы $t_1, \dots, t_n$, в свою очередь, играют роль поддеревьев, которые связаны с дополнительными портами агента $\alpha$.
Если имена входят дважды, то соответствующие листья дерева будут связаны дополнительными дугами.
Заметим, что терм  $\alpha(\vec t)$ не может содержать активных пар.

\item[Уравнения.] Пусть $t$ и $u$~---~некоторые термы.
Тогда неупорядоченную пару $t = u$ будем называть \textit{уравнением}.
С помощью $\Delta, \Theta, \dots$ обозначим мультимножества (т.~е. множества, допускающие многократное вхождение элементов) уравнений.
Примерами уравнений служат $x = \alpha(\vec t)$, $x = y$, $\alpha(\vec t) = \beta(\vec u)$.
Они позволят нам представлять сети с активными парами.

\item[Правила.] \textit{Правилами} будем называть неупорядоченные пары вида
$$
\alpha[t_1, \dots, t_m] \bowtie \beta[u_1, \dots, u_n],
$$
где $(\alpha, \beta) \in \Sigma \times \Sigma$, $m = \text{Ar}(\alpha)$, $n = \text{Ar}(\beta)$, а $t_i$ и $u_i$~---~некоторые термы, причем каждое имя в правиле может иметь только два вхождения (свободные имена в правилах запрещены). 
Неупорядоченная пара $(\alpha, \beta)$ является \textit{активной парой} этого правила (она соответствует левой части правила взаимодействия).
\end{description}

\begin{definition}[имена, входящие в терм]
Обозначим через $\mathcal N(t)$ множество имен, входящих в терм $t$, которое определяется следующим образом:
\begin{align*}
\mathcal N(x) &= \{x\}; \\
\mathcal N(\alpha(t_1, \dots t_n)) &= \mathcal N(t_1) \cup \dots \cup \mathcal N(t_n).
\end{align*}
Данное определение тривиальным образом распространяется на правила и мультимножества уравнений.
\end{definition}

В любом терме можно заменить свободное вхождение одного имени другим, если только сохраняется линейность терма.

\begin{definition}[переименование]
\textit{Переименование} $t[x := y]$ есть результат замены в терме $t$ свободного вхождения имени $x$ новым именем $y$.
Данная операция тривиальным образом распространяется на уравнения и мультимножества уравнений.
\end{definition}

Теперь обобщим переименование до \textit{подстановки}, которая заменяла бы свободное вхождение имени в терме другим термом, как и раньше, сохраняя линейное свойство.

\begin{definition}[подстановка]
\textit{Подстановка} $t[x := u]$ есть результат замены в терме $t$ свободного вхождения имени $x$ термом $u$, если при такой замене сохраняется линейность термов.
\end{definition}

Заметим, что переименование является частным случаем подстановки.
При этом подстановка обладает следующими двумя полезными свойствами.

\begin{theorem}
Пусть $x \not\in \mathcal N(w)$.

\begin{enumerate}
\item Если $y \in \mathcal N(u)$, то $t[x := u][y := w] = t[x := u[y := w]]$.
\item Если $y \not\in \mathcal N(u)$, то $t[x := u][y := w] = t[y := w][x := u]$.
\end{enumerate}
\end{theorem}

Итак, мы ввели достаточно определений, обозначений и свойств, чтобы теперь рассматривать сети в исчислении взаимодействия.

\begin{definition}[конфигурации]
\textit{Конфигурация}~---~это пара $c = (\mathcal R, \langle\vec t\ |\ \Delta\rangle)$, где $\mathcal R$~---~множество правил, $\vec t$~---~последовательность термов $t_1, \dots, t_n$, а $\Delta$~---~мультимножество уравнений.
Каждое имя может иметь не более двух вхождений в $c$.
Если имя имеет одно вхождение, то оно называется \textit{свободным}; в противном случае оно считается \textit{связанным}.
Для простоты иногда будем опускать $\mathcal R$, если из контекста понятно, о каком именно множестве правил взаимодействия идет речь.
Ниже мы обычно будем обозначать конфигурации как $c$ или $c'$.
Последовательность термов $\vec t$ будем называть \textit{головой} конфигурации.
\end{definition}

По существу $\langle\vec t\ |\ \Delta\rangle$ представляет собой сеть взаимодействия, которую мы можем редуцировать, используя правила из $\mathcal R$, а $\Delta$ содержит переименования и активные пары в сети.
Свободные вхождения имен и голова конфигурации вместе формируют интерфейс сети.
Как и в $\lambda$-исчислении, мы будем считать равными конфигурации с точностью до $\alpha$-конверсии, т.~е. замены связанных имен при сохранении линейного свойства.
Действительно, конфигурации, отличающиеся лишь выбором связанных имен, соответствуют одним и тем же сетям взаимодействия.

Преобразовать произвольную сеть взаимодействия в конфигурацию можно, в частности, следующим способом.
Сначала выделим в сети все деревья и направим их главные порты в одном направлении.
Каждую пару деревьев, которые связаны главными портами, представим как уравнение, а деревья, чьи главные порты свободны, добавим в голову конфигурации.
Чтобы проиллюстрировать такое преобразование, ниже мы  приводим простой пример.

\begin{example}
\label{encode}
Операция сложения для натуральных чисел была представлена в сетях взаимодействия (см. параграф~\ref{numbers}) с помощью множества символов $\Sigma = \{0, S, \text{Add}\}$, где $\text{Ar}(0) = 0$, $\text{Ar}(S) = 1$, $\text{Ar}(\text{Add}) = 2$.
На следующей схеме показана сеть, соответствующая выражению $1 + 0$, а затем мы выделяем в ней два дерева и направляем их главными портами вниз.
$$
\begin{tikzpicture}[baseline=(a)]
\matrix[column sep=1em]{
\inetcell(s){$S$}[R] &
\inetcell(a){$\text{Add}$}[L] \\ };
\inetcell[right=of a.above left pax](z){$0$}[L]
\inetcell[left=of s.above middle pax](0){$0$}[R]
\inetwirefree(a.right pax)
\inetwire(s.pal)(a.pal)
\inetwire(s.middle pax)(0.pal)
\inetwire(a.left pax)(z.pal)
\end{tikzpicture}
=
\begin{tikzpicture}[baseline=(a)]
\matrix[column sep=1em]{
\inetcell(a){$\text{Add}$}[D] &
\inetcell(s){$S$}[D] \\ };
\node (x) [above=of a.above right pax] {$x$};
\inetcell[above=of s.above middle pax](0){$0$}[D]
\inetcell[above=of a.above left pax](z){$0$}[D]
\inetwirefree(a.right pax)
\inetwire(s.pal)(a.pal)
\inetwire(s.middle pax)(0.pal)
\inetwire(a.left pax)(z.pal)
\end{tikzpicture}
$$
Теперь несложно получить конфигурацию $\langle x\ |\ \text{Add}(0, x) = S(0)\rangle$, где единственный порт интерфейса помечен $x$, который мы добавили в голову конфигурации.
\end{example}

Обратное преобразование проще: следует построить деревья, соответствующие всем термам в конфигурации, связать порты, соответствующие одинаковым именам, и связать главные порты деревьев, участвующих в уравнениях.

\begin{definition}[редукция]
\label{red}
Отношение \textit{редукции} <<$\rightarrow$>> на конфигурациях порождается следующими тремя правилами.
\begin{description}
\item[Взаимодействие.] Если $\alpha[u'_1, \dots, u'_m] \bowtie \beta[w'_1, \dots, w'_n] \in \mathcal R$ и имеется конфигурация $c = \langle\vec t\ |\ \alpha(u_1, \dots, u_m) = \beta(w_1, \dots, w_n), \Delta\rangle$, то
$$
c \rightarrow \langle\vec t\ |\ u_1 = u'_1, \dots, u_m = u'_m, w_1 = w'_1, \dots, w_n = w'_n, \Delta\rangle.
$$

\item[Разыменование.] Если $x \in \mathcal N(u)$, то
$$
\langle\vec t\ |\ x = t, u = w, \Delta\rangle \rightarrow \langle\vec t\ |\ u[x := t] = w, \Delta\rangle.
$$

\item[Стягивание.] Если $x \in \mathcal N(\vec t)$, то
$$
\langle\vec t\ |\ x = u, \Delta\rangle \rightarrow \langle\vec t[x := u]\ |\ \Delta\rangle.
$$
\end{description}
Рефлексивное транзитивное замыкание отношения редукции обозначим <<$\rightarrow^*$>>.
\end{definition}

Редукция в исчислении взаимодействия демонстрирует фактические затраты на реализацию одного шага взаимодействия, который, как мы видим, заключается не только в построении сети, соответствующей правой части правила взаимодействия, но также разыменовании и стягивании связей.
То же самое происходит и при работе с графическим представлением сетей взаимодействия, хотя обычно взаимодействие кажется атомарной операцией.

\begin{example}[натуральные числа]
\label{calcnat}
Натуральные числа и операция сложения ниже представлены двумя различными способами с использованием исчисления взаимодействия.
Первый из них~---~стандартное представление, тогда как второй вариант эффективнее, так как выполняет сложение двух чисел за фиксированное время $O(1)$.

\begin{enumerate}
\item Пусть $\Sigma = \{0, S, \text{Add}\}$ при $\text{Ar}(0) = 0$, $\text{Ar}(S) = 1$, $\text{Ar}(\text{Add}) = 2$ и множество $\mathcal R$ содержит следующие два правила взаимодействия:
\begin{align*}
\text{Add}[x, S(y)] &\bowtie S[\text{Add}(x, y)]; \\
\text{Add}[x, x] &\bowtie 0.
\end{align*}
Как было показано в примере~\ref{encode}, выражение $1 + 0$ соответствует конфигурации $(\mathcal R, \langle a\ |\ \text{Add}(0, a) = S(0)\rangle)$.
Одна из возможных последовательностей редукций для этой конфигурации выглядит следующим образом:
\begin{align*}
&\langle a\ |\ \text{Add}(0, a) = S(0)\rangle \\
&\rightarrow \langle a\ |\ 0 = x, a = S(y), 0 = \text{Add}(x, y)\rangle \\
&\rightarrow^* \langle S(y)\ |\ 0 = \text{Add}(0, y)\rangle \\
&\rightarrow \langle S(y)\ |\ 0 = x, y = x\rangle \\
&\rightarrow^* \langle S(0)\ |\ \varnothing\rangle.
\end{align*}

\item Пусть $\Sigma = \{S, C, C^*, \text{Add}\}$, причем $\text{Ar}(S) = 1$, $\text{Ar}(C) = \text{Ar}(C^*) = \text{Ar}(\text{Add}) = 2$.
Натуральное число представим в виде разностного списка агентов $S$, так, что агент $C$ будет иметь связи с началом и концом такого списка.
Тогда нуль $0$ будет соответствовать конфигурации $\langle C(x, x)\ |\ \varnothing\rangle$, а произвольное положительное число $n$ примет форму $\langle C(x, S^n(x))\ |\ \varnothing\rangle$.

Множество $\mathcal R$ будет содержать следующие два правила взаимодействия:
\begin{align*}
C[x, y] &\bowtie \text{Add}[C^*(z, y), C(x, z)];\\
C[x, y] &\bowtie C^*[y, x].
\end{align*}
Приведем в качестве примера выражение $m + n$, чтобы показать, что сложение выполняется за время $O(1)$ вне зависимости от самих чисел $m$ и $n$:
\begin{align*}
&\langle z\ |\ C(x, S^m(x)) = \text{Add}(C(y, S^n(y)), z)\rangle \\
&\rightarrow \langle z\ |\ x = x', S^m(x) = y', C(y, S^n(y)) = C^*(z', y'), z = C(x', z')\rangle \\
&\rightarrow^* \langle C(x, z')\ |\  C(y, S^n(y)) = C^*(z', S^m(x))\rangle \\
&\rightarrow \langle C(x, z')\ |\  y = x', S^n(y) = y', z' = y', S^m(x) = x'\rangle \\
&\rightarrow^* \langle C(x, z')\ |\  S^n(S^m(x)) = z'\rangle \\
&\rightarrow \langle C(x, S^{m + n}(x))\ |\ \varnothing\rangle.
\end{align*}
\end{enumerate}
\end{example}

Исчисление взаимодействия~---~Тьюринг-полная модель вычислений.
Стало быть, в нем имеет место и проблема останова (т.~е. в общем случае нельзя определить, будет ли конечной последовательность редукций).
В следующем примере рассматривается конфигурация, которая приводит к бесконечной последовательности редукций.

\begin{example}[бесконечный цикл]
\label{endless}
Рассмотрим конфигурацию $\langle x, y\ |\ \alpha(x) = \beta(\alpha(y))\rangle$ с правилом взаимодействия $\alpha[x] \bowtie \beta[\beta(\alpha(x))]$.
Тогда мы имеем последовательность редукций, которая оказывается бесконечной:
\begin{align*}
&\langle x, y\ |\ \alpha(x) = \beta(\alpha(y))\rangle \\
&\rightarrow \langle x, y\ |\ x = x', \alpha(y) = \beta(\alpha(x'))\rangle \\
&\rightarrow \langle x', y\ |\ \alpha(y) = \beta(\alpha(x'))\rangle \\
&\rightarrow \dots
\end{align*}
\end{example}

Обратим внимание, что исчисление взаимодействия может представлять системы взаимодействия с двумя ограничениями.

Во-первых, в исчислении взаимодействия нельзя записать правило с активной парой в правой части.
Впрочем, это ограничение не является проблемой, так как несложно показать, что класс систем взаимодействия без активных пар в правой части правил имеет ту же мощность с вычислительной точки зрения.

Во-вторых, любое правило без интерфейса неизбежно имеет пустую сеть в правой части.
Однако, в дальнейшем нас не будут интересовать несвязные части сети, так как результат вычисления определяется только относительно интерфейса.

\subsection{Свойства исчисления взаимодействия}

Данный параграф посвящен обсуждению свойств редукции, которая порождается правилами взаимодействия, разыменования и стягивания (см. определение~\ref{red}).
Мы уже рассматривали свойства систем взаимодействия в их графической форме.
Все те же свойства справедливы и для исчисления взаимодействия.

\begin{theorem}[конфлюэнтность]
Отношение редукции <<$\rightarrow$>> на конфигурациях обладает свойством сильной конфлюэнтности: если $c \rightarrow d$ и $c \rightarrow e$, где $d$ и $e$~---~различные конфигурации, то существует конфигурация $c'$, такая, что $d \rightarrow c'$ и $e \rightarrow c'$.
\end{theorem}

Пусть $c$~---~конфигурация.
Если не существует конфигурации $c'$, такой, что $c \rightarrow c'$, то $c$ называется \textit{нормальной формой}.

Пусть теперь $c$~---~некоторая конфигурация, а $c'$~---~нормальная форма.
Тогда $c \downarrow c'$, если $c \rightarrow^* c'$.
Из свойства сильной конфлюэнтности непосредственно следует единственность нормальной формы, если она существует: если $c \downarrow d$ и $c \downarrow e$, то $d = e$.

Заметим, что проблема останова (см. пример~\ref{endless}) возникает лишь из-за правила взаимодействия, но не правил разыменования и стягивания.
Разыменование и стягивание порождают только конечные последовательсти редукций, так как они всегда уменьшают число уравнений в конфигурации.

\subsection{Нормальные формы и стратегии}

Хотя мы неоднократно подчеркивали свойство сильной конфлюэнтности систем взаимодействия, в общем случае существуют различные последовательности редукций, приводящие к нормальной форме, если она существует.
Более того, существуют и другие понятия нормальных форм, в частности для так называемых <<ленивых>> вычислений.

Ниже мы определим более слабое понятие нормальной формы, которая является аналогом головной нормальной формы в $\lambda$-исчислении.
Последние часто используются в реализациях функциональных языков программирования.

\begin{definition}[головная нормальная форма]
Конфигурация $(\mathcal R, \langle\vec t\ |\ \Delta\rangle)$ есть \textit{головная нормальная форма}, если каждый $t_i$ в $\vec t$ имеет одну из следующих форм.
\begin{itemize}
\item $\alpha(\vec s)$.
Например, $\langle S(x)\ |\ x = 0\rangle$.
\item $x$, где $x \in \mathcal N(t_j)$ для $j \neq i$.
При этом связь, соответствующую имени $x$, будем называть \textit{открытой}.
Например, $\langle x, x\ |\ \Delta\rangle$.
\item $x$, если мультимножество $\Delta$ содержит уравнение вида $y = s$, где $s$~---~терм, причем $x \in \mathcal N(s)$ и $y \in \mathcal N(s)$.
В этом случае уравнение $y = s$ и соответствующую этому уравнению часть сети называют \textit{циклическим деревом}.
Например, конфигурация $\langle x\ |\ y = \alpha(\beta(y), x), \Delta\rangle$ содержит циклическое дерево $y = \alpha(\beta(y), x)$.
\end{itemize}
\end{definition}

Иными словами, конфигурация находится в головной нормальной форме, если ее голова представляет собой интерфейс, каждый свободный порт которого связан либо с главным портом, либо с другим портом этого интерфейса, либо с дополнительным портом, который никогда не станет главным после последовательности редукций.

Проиллюстрируем понятие головной нормальной формы на нескольких схемах.
На первой~---~часть сети, соответствующей конфигурации $\langle\alpha(t_1, \dots, t_n)\ |\ \Delta\rangle$, которая состоит из агента $\alpha$ со свободным главным портом и термов $t_i$, связанных с дополнительными портами $\alpha$.
На второй схеме показана открытая связь (через агент $\delta$).
$$
\begin{tikzpicture}[baseline=(0.base)]
\inetcell(a){$\phantom X\alpha\phantom X$}[U]
\inetwirefree(a.pal)
\inetwirefree(a.left pax)
\inetwirefree(a.right pax)
\node (0) [above=of a.above pal] {$\langle\alpha(t_1, \dots, t_n)\ |\ \Delta\rangle$};
\node (1) [below=of a.above left pax] {$t_1$};
\node (i) [below=of a.above middle pax] {$\dots$};
\node (n) [below=of a.above right pax] {$t_n$};
\end{tikzpicture}
\qquad
\begin{tikzpicture}[baseline=(0.base)]
\inetcell(a){$\phantom x\delta\phantom x$}[U]
\inetcell[below=of a.above right pax](z){$0$}[U]
\inetwirefree(a.pal)
\inetwirefree(a.left pax)
\inetwire(z.pal)(a.right pax)
\node (0) [above=of a.above pal] {$\langle x, \delta(x, 0)\ |\ \varnothing\rangle$};
\node (1) [below=of a.above left pax] {$x$};
\end{tikzpicture}
$$

Следующие две схемы являются примерами конфигураций, которые содержат циклические деревья: уравнения $\alpha(x, \beta(y)) = y$ и $\delta(S(x), y) = y$.
$$
\begin{tikzpicture}[baseline=(c.base)]
\matrix[column sep=1em]{
\inetcell(a){$\alpha$}[U] &
\inetcell(b){$\beta$}[D] \\ };
\node (x) [below=of a.above left pax] {$x$};
\node (c) [above=of a.above pal] {$\langle x\ |\ \alpha(x, \beta(y)) = y\rangle$};
\inetwirefree(a.left pax)
\inetwire(a.pal)(b.middle pax)
\inetwire(b.pal)(a.right pax)
\end{tikzpicture}
\qquad
\begin{tikzpicture}[baseline=(c.base)]
\matrix[column sep=1em]{
\inetcell(a){$\delta$}[U] &
\inetcell[opacity=0](b){$\beta$}[D] \\ };
\inetcell[below=of a.above left pax](s){$S$}[U]
\node (x) [below=of s.above middle pax] {$x$};
\node (c) [above=of a.above pal] {$\langle x\ |\ \delta(S(x), y) = y\rangle$};
\inetwirecoords(b.pal)(b.middle pax)
\inetwirefree(s.middle pax)
\inetwire(a.left pax)(s.pal)
\inetwire(a.pal)(b.middle pax)
\inetwire(b.pal)(a.right pax)
\end{tikzpicture}
$$

\section{Недетерминированные расширения}

Системы взаимодействия можно считать моделью распределенных вычислений в том смысле, что правила взаимодействия применимы в любом порядке или даже одновременно независимо друг от друга в любых частях сети (синхронизация при этом не требуется благодаря свойству сильной конфлюэнтности у редукции).
Но сети взаимодействия сами по себе не способны представлять недетерминированные вычисления, которые характерны для параллельных вычислительных систем.

Были предложены несколько расширений понятия о системах взаимодействия, чтобы стало возможным моделировать внутри них недетерминированный выбор.
В частности, мы могли бы допустить несколько правил взаимодействия для одной и той же активной пары, причем так, чтобы выбор одного из этих правил происходил произвольным образом.
Но сети взаимодействия не содержат выразительных средств для полноценной модели недетерминированных вычислений.

Если же обобщить понятие агента так, чтобы взаимодействие могло иметь место на двух и более портах (т.~е. допустить агенты с несколькими главными портами), то полученная система уже будет обладать необходимой выразительностью.
На самом деле, достаточно иметь всего один агент с двумя главными портами.
Он и будет представлять собой неоднозначный выбор.
Обычно такой агент обозначается $\text{Amb}$.
Правила взаимодействия для этого агента выглядят следующим образом.
$$
\begin{tikzpicture}[baseline=(b.base)]
\matrix[row sep=2em]{
\inetcell(a){$\text{Amb}$}[D] \\
\inetcell[opacity=0](p){$X$}[D] \\ };
\node (m) [above=of a.above right pax] {$m$};
\node (t) [above=of a.above left pax] {$a$};
\node (b) [below=of p.left pax] {$b$};
\inetcell[below=of p.right pax](x){$\phantom x\alpha\phantom x$}[U]
\node (d) [below=of x.middle pax] {$\dots$};
\inetwirefree(a.left pax)
\inetwirefree(a.right pax)
\inetwirefree(x.left pax)
\inetwirefree(x.right pax)
\inetwire(a.pal)(p.left pax)
\inetwire(a.pal)(x.pal)
\end{tikzpicture}
\rightarrow
\begin{tikzpicture}[baseline=(x)]
\matrix[row sep=1em]{
\node (m) {$m$}; & \node (t) {$a$}; \\
\inetcell(x){$\phantom x\alpha\phantom x$}[U] & \\
& \node (b) {$b$}; \\ };
\node (d) [below=of x.middle pax] {$\dots$};
\inetwirefree(x.left pax)
\inetwirefree(x.right pax)
\inetwirecoords(x.pal)(m)
\inetwirecoords(t)(b)
\end{tikzpicture}
\qquad
\begin{tikzpicture}[baseline=(b.base)]
\matrix[row sep=2em]{
\inetcell(a){$\text{Amb}$}[D] \\
\inetcell[opacity=0](p){$X$}[D] \\ };
\node (m) [above=of a.above right pax] {$m$};
\node (t) [above=of a.above left pax] {$a$};
\node (b) [below=of p.right pax] {$b$};
\inetcell[below=of p.left pax](x){$\phantom x\alpha\phantom x$}[U]
\node (d) [below=of x.middle pax] {$\dots$};
\inetwirefree(a.left pax)
\inetwirefree(a.right pax)
\inetwirefree(x.left pax)
\inetwirefree(x.right pax)
\inetwire(a.pal)(p.right pax)
\inetwire(a.pal)(x.pal)
\end{tikzpicture}
\rightarrow
\begin{tikzpicture}[baseline=(x)]
\matrix[row sep=1em]{
\node (t) {$a$}; & \node (m) {$m$}; \\
& \inetcell(x){$\phantom x\alpha\phantom x$}[U] \\
\node (b) {$b$}; & \\ };
\node (d) [below=of x.middle pax] {$\dots$};
\inetwirefree(x.left pax)
\inetwirefree(x.right pax)
\inetwirecoords(x.pal)(m)
\inetwirecoords(t)(b)
\end{tikzpicture}
$$

Когда агент $\alpha$ связан своим главным портом с одним из главных портов $\text{Amb}$, может иметь место взаимодействие, в результате которого $\alpha$ оказывается связанным с \textit{основным} свободным портом $m$ в интерфейсе на стороне агента $\text{Amb}$.
Если оба главных порта $\text{Amb}$ связаны с главными портами других агентов, то выбор одного из правил взаимодействия выше недетерминирован.

Воспользуемся агентом $\text{Amb}$, чтобы представить функцию <<параллельное или>> $\text{ParallelOr}$~---~это необычная логическая операция, которая возвращает <<истину>> $\text{True}$, если один из ее двух аргументов равен $\text{True}$, причем вне зависимости от того, вычислен ли другой аргумент.

\begin{example}
Функция $\text{ParallelOr}$ должна немедленно возвращать результат $\text{True}$, как только один из ее аргументов принял значение $\text{True}$, даже когда другой из ее аргументов вовсе не определен.
С помощью агента $\text{Amb}$ мы можем представить эту функцию в виде сети, изображенной на следующей схеме.
$$
\begin{tikzpicture}
\matrix[column sep=2em]{
\inetcell(o){$\text{Or}$}[D] &
\inetcell(a){$\text{Amb}$}[R] &
\inetcell[opacity=0](p){$X$}[R] \\ };
\inetwirefree(o.right pax)
\inetwire(o.pal)(a.right pax)
\inetwire(o.left pax)(a.left pax)
\inetwire(a.pal)(p.left pax)
\inetwire(a.pal)(p.right pax)
\end{tikzpicture}
$$
Здесь $\text{Or}$ представляет собой обычное логическое <<или>>, определяемое следующими двумя правилами взаимодействия.
$$
\begin{tikzpicture}[baseline=(p.base)]
\matrix{
\inetcell(t){$T$}[R] &
\node (p) {$\phantom x$}; &
\inetcell(o){$\text{Or}$}[L] \\ };
\inetwirefree(o.left pax)
\inetwirefree(o.right pax)
\inetwire(o.pal)(t.pal)
\end{tikzpicture}
\rightarrow
\begin{tikzpicture}[baseline=(p.base)]
\matrix{
\inetcell(t){$T$}[U] &
\node (p) {$\phantom x$}; &
\inetcell(e){$\epsilon$}[D] \\ };
\inetwirefree(t.pal)
\inetwirefree(e.pal)
\end{tikzpicture}
\qquad
\begin{tikzpicture}[baseline=(p.base)]
\matrix{
\inetcell(f){$F$}[R] &
\node (p) {$\phantom x$}; &
\inetcell(o){$\text{Or}$}[L] \\ };
\inetwirefree(o.left pax)
\inetwirefree(o.right pax)
\inetwire(o.pal)(f.pal)
\end{tikzpicture}
\rightarrow
\begin{tikzpicture}[baseline=(p.base)]
\matrix[row sep=1em]{
\node (t) {$\phantom x$}; \\
\node (p) {$\phantom x$}; \\
\node (b) {$\phantom x$}; \\ };
\inetwirecoords(t)(b)
\end{tikzpicture}
$$
\end{example}

Модель вычислений, соответствующая сетям взаимодействия с дополнительным агентом $\text{Amb}$ обладает в определенном смысле большей мощностью, чем обычные системы взаимодействия, позволяя представить недетерминированные вычисления и параллельные функции, такие как <<параллельное или>> $\text{ParallelOr}$.

\section{Дополнительная литература}

Оригинальная статья~\cite{inet} вводит понятие сетей взаимодействия и содержит много примеров их использования.
Больше информации о комбинаторах взаимодействия, а также доказательство их универсальности читатель может получить в~\cite{icomb}.
В~\cite{impl} обсуждаются реализации сетей взаимодействия.
Описанная выше текстовая запись для сетей взаимодействия впервые была введена в оригинальной статье~\cite{inet}, в то время как исчисление, основанное на этой нотации, определяется в~\cite{calc}.
Также~\cite{calc} содержит другие интересные результаты о нормальных формах и стратегиях.

\section{Упражнения}
\label{exercise}

\begin{enumerate}
\item В сетях взаимодействия, определить следующие арифметические операции для натуральных чисел, представленных нулем $0$ и функцией следования $S$:
\begin{itemize}
\item проверка на нуль $\text{Zero}$, которая приводит к результату <<истина>> $\text{True}$, если число равно нулю $0$, и к результату <<ложь>> $\text{False}$~---~в противном случае;
\item минимум $\text{Min}$, которая вычисляет наименьшее из двух чисел;
\item факториал $\text{Fact}$ произвольного натурального числа.
\end{itemize}

\item Построить сеть взаимодействия, приводящую к бесконечному циклу.

\item Дополнить определение системы взаимодействия для комбинаторной логики в параграфе~\ref{complete}.
Точнее, определите наборы агентов и правил, необходимые, чтобы представить комбинатор $S$ (достаточно три агента и три правила).

\item
\begin{enumerate}
\item Построить систему взаимодействия для вычисления логического <<и>> $\text{And}$.
\item Изобразить сеть взаимодействия, представляющую выражение
$$
(\text{True}\ \text{And}\ \text{False})\ \text{And}\ \text{True}.
$$
Сколько шагов потребуется, чтобы получить нормальную форму?
\item Модифицировать систему так, чтобы результат был $\text{True}$ тогда и только тогда, когда оба аргумента имеют одинаковое значение (т.~е. либо оба $\text{True}$, либо оба $\text{False}$).
\end{enumerate}

\item Представить функцию, которая, для двух данных списков $l_1$ и $l_2$, строит список, содержащий элементы $l_1$, чередующиеся элементами $l_2$.
Например, результатом чередования списков $[0, 2, 4]$ и $[1, 3]$ является список $[0, 1, 2, 3, 4]$.

\item В примере~\ref{calcnat} представлена текстовая запись правил для сложения чисел.
Как аналогичным образом записать правила для умножения чисел из параграфа~\ref{numbers}?

\item Почему сети взаимодействия в чистом виде не могут применяться в качестве модели нетерминированных вычислений?

\item Определить функцию <<параллельное и>> $\text{ParallelAnd}$ с помощью агента $\text{Amb}$.
Функция $\text{ParallelAnd}$ отличается от обычной тем, что немедленно возвращает <<ложь>> $\text{False}$, как только один из аргументов принимает значение $\text{False}$.
\end{enumerate}

\appendix
\section{Ответы к некоторым упражнениям}

\begin{enumerate}
\addtocounter{enumi}{3}
\item
\begin{enumerate}
\item
$$
\begin{tikzpicture}[baseline=(p.base)]
\matrix[column sep=1em]{
\inetcell(F){$F$}[R] &
\inetcell(And){$\text{And}$}[L] \\ };
\inetwirefree(And.left pax)
\inetwirefree(And.right pax)
\inetwire(F.pal)(And.pal)
\node (p) [right=of And.pal] {$\phantom{p}$};
\node (m) [right=of And.above right pax] {$m$};
\node (x) [right=of And.above left pax] {$x$};
\end{tikzpicture}
\rightarrow
\begin{tikzpicture}[baseline=(p.base)]
\matrix{
\inetcell(F){$F$}[R] \\
\node (p) {$\phantom{p}$}; \\
\inetcell(e){$\epsilon$}[R] \\ };
\inetwirefree(F.pal)
\inetwirefree(e.pal)
\node (m) [right=of F.above pal] {$m$};
\node (x) [right=of e.above pal] {$x$};
\end{tikzpicture}
\qquad
\begin{tikzpicture}[baseline=(p.base)]
\matrix[column sep=1em]{
\inetcell(T){$T$}[R] &
\inetcell(And){$\text{And}$}[L] \\ };
\inetwirefree(And.left pax)
\inetwirefree(And.right pax)
\inetwire(T.pal)(And.pal)
\node (p) [right=of And.pal] {$\phantom{p}$};
\node (m) [right=of And.above right pax] {$m$};
\node (x) [right=of And.above left pax] {$x$};
\end{tikzpicture}
\rightarrow
\begin{tikzpicture}[baseline=(p.base)]
\matrix{
\node (m) {$m$}; \\
\node (p) {$\phantom{p}$}; \\
\node (x) {$x$}; \\ };
\inetwirecoords(m)(x)
\end{tikzpicture}
$$

\item Обойдемся без иллюстрации.
Cеть содержит активную пару $T \bowtie \text{And}$, которая порождает $F \bowtie \text{And}$.
Это взаимодействие приводит к созданию нового агента $F$ и активной пары $T \bowtie \epsilon$.
Через три шага получаем $\text{False}$.
\end{enumerate}

\addtocounter{enumi}{1}
\item
\begin{align*}
\text{Mult}[\epsilon, 0] &\bowtie 0; \\
\text{Mult}[\delta(x, y), z] &\bowtie S[\text{Mult}(y, \text{Add}(x, z))].
\end{align*}

\item Сети взаимодействия детерминированны по своей сути, обладая свойством так называемой сильной конфлюэнтности.
Из этого свойства следует, что любые последовательности редукций приводят сеть к одному и тому же состоянию.
\end{enumerate}

\end{document}